\documentclass[reprint,superscriptaddress,amsmath,amssymb,aps,pra]{revtex4-2}

\usepackage{graphicx}
\usepackage{dcolumn}
\usepackage{bm}
\usepackage{float}
\usepackage[colorlinks=true, allcolors=blue]{hyperref}

\begin{document}

\preprint{APS/123-QED}

\title{Manipulating the transverse spin angular momentum and Belinfante momentum of spin-polarized light by a tilted stratified medium in optical tweezers}

\author{Sauvik Roy}
\affiliation{Department of Physical Sciences, IISER-Kolkata, Mohanpur 741246, India}

\author{Nirmalya Ghosh}
\email{nghosh@iiserkol.ac.in}
\affiliation{Department of Physical Sciences, IISER-Kolkata, Mohanpur 741246, India}

\author{Ayan Banerjee}
\email{ayan@iiserkol.ac.in}
\affiliation{Department of Physical Sciences, IISER-Kolkata, Mohanpur 741246, India}

\author{Subhasish Dutta Gupta}
\email{sdghyderabad@gmail.com}
\affiliation{Department of Physical Sciences, IISER-Kolkata, Mohanpur 741246, India}
\affiliation{School of Physics, Hyderabad Central University, India}
\affiliation{Tata Institute of Fundamental Research Hyderabad, India}

\date{\today}

\begin{abstract}
	In the recent past, optical tweezers incorporating a stratified medium have been exploited to generate complex translational and rotational dynamics in mesoscopic particles due to the coupling between the spin and orbital angular momentum  of the light, generated as a consequence of the tight focusing of light by a high numerical aperture objective lens into the stratified medium. Here, we consider an optical tweezers system with a tilted stratified medium (direction of stratification at an angle with the axis of the incident beam), and show that for input circularly polarized Gaussian beams, the resulting spin-orbit interaction deeply influences the generation of transverse spin angular momentum (TSAM) and Belinfante momentum of light, and allows additional control on their magnitude.  Importantly, the TSAM generated in our system consists of both the orthogonal components, which is in sharp contrast to the case of evanescent waves and surface plasmons, where only one of the TSAM components are generated. The broken symmetry due to the tilt ensures that, depending upon the helicity of the input beam, the magnitude of the mutually orthogonal components of the TSAM and the Belinfante momentum depend entirely on the tilt angle. This may prove to be an effective handle in exotic spin-controlled manipulation of particles in experiments.
	
\end{abstract}

\maketitle


\section{Introduction}

It has been known for decades that in addition to linear momentum, a light wave can carry both  spin and orbital angular momentum (SAM and OAM, respectively) \cite{bliokh2014extraordinary,berry2009optical}. While the SAM of light is associated with the polarization degrees of freedom of the electromagnetic wave, OAM is related to the evolution of the wave vector and the phase structure of the light beam. Although the presence of such SAM and OAM in any electromagnetic field is general in nature, their manifestations for an arbitrary field configurations is rather intertwined and complex. For a circularly or elliptically polarized plane wave or a paraxial Gaussian beam, the intrinsic SAM is determined by the helicity $\sigma (-1 \leq \sigma \leq 1)$ of the electromagnetic field. However, higher order paraxial beams, having complex amplitude and phase distributions can carry both intrinsic and extrinsic OAM \cite{PhysRevLett.88.053601,saha2018transverse}. Spin-orbit interaction (SOI) of light dealing with the interactions and inter-conversions between intrinsic SAM, and intrinsic and extrinsic OAM degrees of freedom of classical light beam has led to a number of striking optical phenomena such as generation of spin-dependent optical vortices, SAM and OAM dependent shift of the trajectory of light beams, the so-called Spin and Orbital Hall effect of light, etc. \cite{bliokh2015quantum,athira2021experimental}.  Such SOI effects have been observed in various optical interactions ranging from reflection/refraction of optical beams at interfaces, tight focusing of fundamental and higher order Gaussian beams, high numerical aperture (NA) imaging geometry, propagation through  inhomogeneous anisotropic media, and so forth. Each of the observed SOI effects are discernable by important fundamental or useful application aspects.
\par
Note that most of the SOI effects observed for propagating light beams in diverse micro- and nano-scale optical systems can be interpreted through the evolution of various types of geometric phases (and its spatial or momentum gradients) which can be understood through the conventional longitudinal (along the direction of the wave vector) angular momentum, and the transverse electric and the magnetic field components of light. These types of SOI effects dealing with the transverse field components have been conveniently modeled using the Debye-Wolf theory for tight focusing, Mie theory for scattering, or the conventional Jones matrix algebra for SOI in inhomogeneous anisotropic media. However, recent studies have demonstrated a different kind of SOI effect that is exclusively related to the transverse component of angular momentum in highly non-paraxial or evanescent fields  \cite{aiello2015transverse,pal2020direct}. It has been recognized that for highly structured fields, a strong longitudinal component of the field leads to the appearance of both a transverse component of SAM, and a spin-dependent transverse momentum component - also known as the Belinfante spin momentum (BSM). Surprisingly, this transverse SAM, initially observed for evanescent fields such as those arising in surface plasmons, was found to be completely independent of the  helicity of the input light wave. Later on, it was observed that such extraordinary transverse SAM (TSAM) is also manifested in other types of structured fields, such as in the scattering of plane waves from micro and nano scale scatterers, tight focusing  of fundamental or higher order Gaussian modes \cite{neugebauer2015measuring, mukherjee2016coherent,singh2018transverse}, etc. This can be understood by recognizing that TSAM appears when the longitudinal component of the field is shifted in phase with respect to the transverse field, which is generated at the scattered near-field, or in the focal plane of a tightly focused beam \cite{bliokh2014extraordinary}. A number of studies have therefore addressed means for enhancing  the TSAM through the enhancement of the longitudinal field component  \cite{saha2016transverse}. This novel type of SOI dealing with the transverse angular momentum (both SAM and OAM) and spin-dependent transverse BSM of light has led to observation of remarkable effects in spin-controlled directional coupling between circularly polarized incident light beams and transversely propagating surface modes in nano-fibers, metal surfaces, waveguides etc. 
\cite{abujetas2020spin,svak2018transverse,saha2018effects,shao2018spin}. This has opened up exciting new opportunities for the development of spin orbit photonic nano devices for control and manipulation of light at nanometer length scales. While various intriguing manifestations of this TSAM has already been studied in a number of optical interactions and systems,  to the best of our knowledge, the dependence of the TSAM and BSM on processes that break geometrical symmetry while the light propagates in an optical medium has not been studied. Such a breaking of symmetry may be simply realized by inserting a refractive index (RI) stratified medium in the path of tightly focused light and tilting the medium with respect to the beam propagation direction. In this paper,  we present an in-depth theoretical analysis and simulations on the generation of TSAM and BSM in such a system, where a tightly focused paraxial Gaussian laser beam is incident on a RI stratified medium obliquely, this breaking the axial symmetry for all input field distributions. The stratification is also important to study since it has important implications in experiments, with the associated spherical aberration deeply affecting the intensity distribution axially so as to facilitate off-axis trapping of particles, and the observation of several interesting and intriguing effects such as the spin Hall shift \cite{roy2014manifestations}, and effects of the BSM \cite{pal2020direct}.
\par
The structure of the paper is as follows. In Section II, we outline the theoretical framework to evaluate the field components near the focus in the stratified medium with and without the tilt. We then proceed to describe the  numerical results for a typical tweezers setup with a high-contrast stratified medium, and analyse and discuss the results in Section III, after which we summarize our findings in the Conclusions section. 

\section{Theoretical framework and mathematical formulation}
There have been extensive studies on the Debye-Wolf formalism \cite{richards1959electromagnetic,wolf1959electromagnetic} and its possible extensions to determine the electromagnetic field components near the focus of an incident Gaussian beam inside a stratified medium with a view to tackle generic experimental situations in optical tweezers \cite{torok1997electromagnetic, haldar2012self}. Such techniques require independent evaluations of 2D vector diffraction integrals for all constituent plane waves (spatial harmonics) for a particular point of interest. An extension of this formalism using a hybrid transfer function can lead to the desired field profiles anywhere inside or after the stratified medium. Most of these studies focus on  systems where the direction of stratification coincides with the optical axis of the high numerical aperture (NA) microscope objective lens which forms an integral component of optical tweezers. However, the calculation of the fields - for the case where the stratified medium is tilted with respect to the symmetry axis of the lens by further modifying the transfer function - becomes nontrivial. When the interfaces are tilted, a ray of light coming out of the exit aperture of the lens does not remain in the same meridional plane because it suffers reflection/refraction at the first interface. So, from the ray optics point of view, there occurs a change in the plane of incidence at the very first interface. However, the ray remains in the new plane for all the successive reflections/refractions. Also, due to the coupled nature of spin and polarization properties in SOI, the change in polarization needs to be continuously tracked for the propagation of all the plane waves. Since this formalism depicts the field as a vectorial superposition of plane waves, focusing and propagation through stratified medium gets decoupled \cite{munro2018tool}, and one can find field components relatively easily. Such decoupling of a single problem into two independent problems allows us to numerically find the optical fields for the tilted interface structure for an incident beam with an arbitrary field profile.  
\par
\begin{figure}[ht]
\centering
\includegraphics[width=0.45\textwidth]{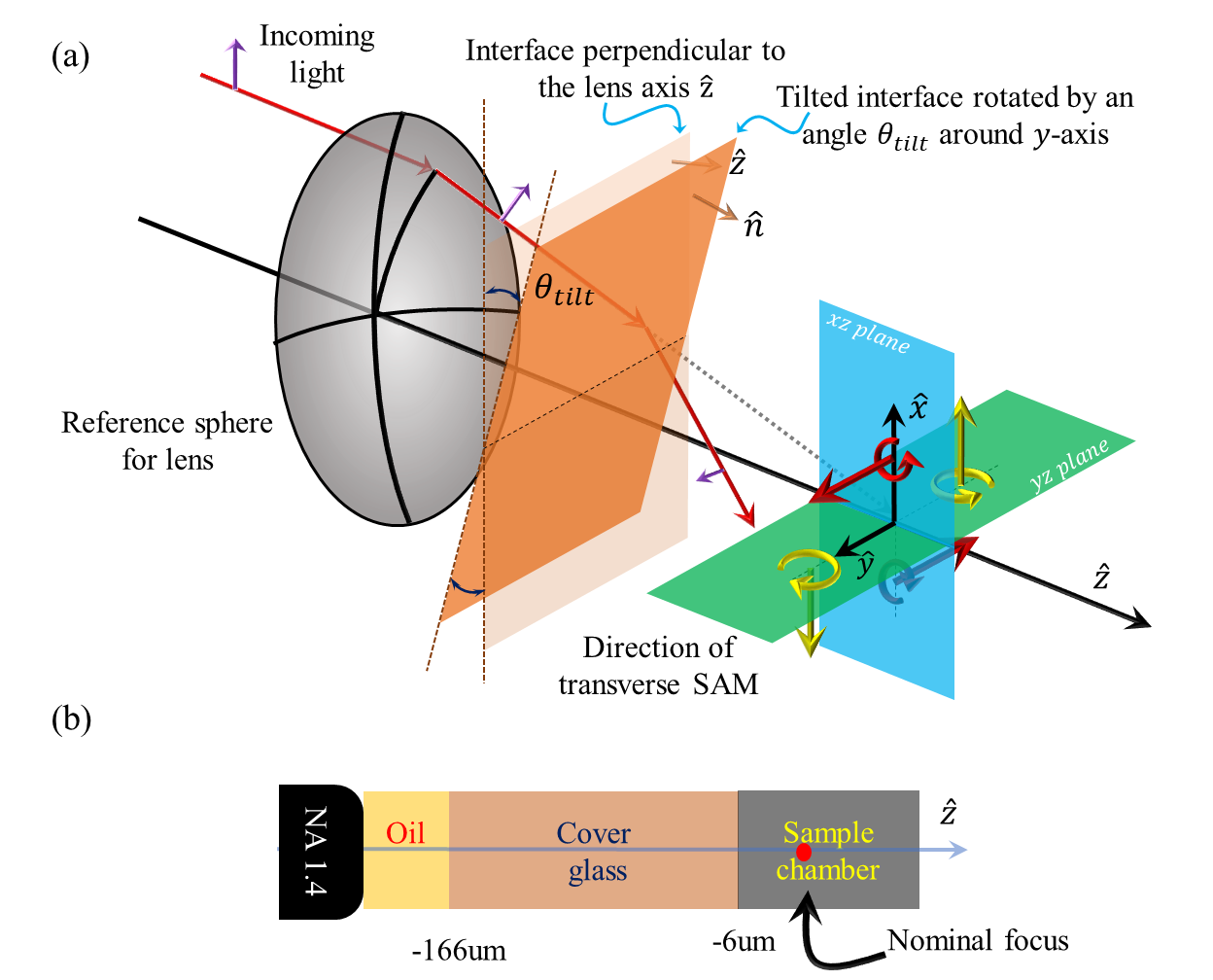}
\caption{(a) Schematic of electromagnetic focusing through tilted stratified medium. Bold red and yellow arrows denote the direction of transverse SAM. (b) Schematic of the optical tweezers setup developed around a stratified medium which we consider in our studies.}
\label{tilted_diagram}
\end{figure}

In what follows, we follow Ref.~\cite{munro2018tool} to present the basic steps and generalize the approach for the case of a tilted stratified medium. We start our theoretical model with a lens system of arbitrary NA and focal length $f$, with its axis along $\bf{\hat{z}}$, as shown in Fig.~\ref{tilted_diagram}. This system focuses an incoming Gaussian beam at its nominal focus $O$. The origin of the coordinate system is considered coincident with the nominal focus. Henceforth we assume the incident fields to be monochromatic with the temporal factor given by $\exp{(-i\omega t)}$.
\subsection{The Debye–Wolf integral and modeling the tilted stratified medium}
The time-independent electric field at a point $p$ in a homogeneous space of RI $n_0$ is
\begin{equation}
        {\bf E}(p)=-ic\iint_{\Omega} \frac{{\boldsymbol{\epsilon}}(s_{x},s_{y} )}{s_{z}} e^{in_{0}k_{0}({\bf s}.\boldsymbol{r}_{p})} \,ds_x\,ds_y
\end{equation}
Here, $c=\frac{n_{0} k_{0} f}{2 \pi}$, ${\boldsymbol{\epsilon}}(s_{x},s_{y})$ is the electric field strength vector on the reference Gaussian sphere, ${\bf s}=(s_{x},s_{y},s_{z}) \equiv ({\bf s}_{\perp},\sqrt{1-|{\bf s}_{\perp}|^2})$ is typically the direction of a spherical wave front, $k_{0}$ is the free space wave number, $\Omega$ is the solid angle formed by the geometric rays converging from the lens.

The determination of the field strength vector ${\boldsymbol{\epsilon}}(s_{x},s_{y})$ on the reference sphere for any arbitrary field, using geometrical optics, involves total three rotations of the coordinate system. These three rotations are arranged to construct a transfer function $A$, in such a way that it correctly mimics a lens action in the lab/global frame. The transfer function looks like
\begin{equation}
    A=R_{z}(-\phi)R_{y}(\theta)R_{z}(\phi).
\end{equation}
Here the $SO(3)$ rotation matrices, i.e., $R_{z}$ and $R_{y}$ take the explicit form as
\begin{equation}
R_z(\phi)=
    \begin{bmatrix}
\cos{\phi} & \sin{\phi} & 0 \\
 -\sin{\phi}& \cos{\phi}& 0 \\
0 & 0 & 1
\end{bmatrix}
\end{equation}
\begin{equation}
    R_y(\theta)=
    \begin{bmatrix}
\cos{\theta} & 0 & -\sin{\theta} \\
0 & 1 &  0 \\
\sin{\theta} & 0 & \cos{\theta}
\end{bmatrix}
\end{equation}
The  generalised Jones vector formalism then connects the injected electric field distribution on the back focal plane (bfp) of the lens to that on the reference sphere by the help of this transfer function.
\begin{equation}
    {\boldsymbol{\epsilon}}(s_{x},s_{y})=\sqrt{\cos{\theta}}A{\boldsymbol{\epsilon}}^{in}_{_{bfp}}.
\end{equation}
%
\begin{equation}\label{theta_phi_dependency}
{\boldsymbol{\epsilon}}(s_{x},s_{y})
=
\begin{bmatrix}
a-b\cos{2\phi} & -b\sin{2\phi} \\
-b\sin{2\phi} & a+b\cos{2\phi} \\
-c\cos{\phi} & -c\sin{\phi}
\end{bmatrix}
\begin{bmatrix}
{\boldsymbol{\epsilon}}^x_{_{bfp}}\\{\boldsymbol{\epsilon}}^y_{_{bfp}}\\
\end{bmatrix}
\end{equation}
Where, ${\boldsymbol{\epsilon}}^{in}_{_{bfp}}=({\boldsymbol{\epsilon}}^x_{_{bfp}}, {\boldsymbol{\epsilon}}^y_{_{bfp}})$ is an arbitrary field in the back focal plane, $a=\frac{1}{2}\sqrt{\cos{\theta}} (1+\cos{\theta})$, $b=\frac{1}{2}\sqrt{\cos{\theta}}(1-\cos{\theta})$, $c=\sqrt{\cos{\theta}}\sin{\theta}$. We emphasize that the input field should not have a component in the longitudinal direction.

In order to incorporate the effects of the stratified medium, we follow a matrix formalism where each plane wave incident from the objective lens is decoupled into TE and TM modes at desired planes transverse to the stratification axis, $\boldsymbol{\xi}$. The electric field at a point $(x,y,z)$ is given by the superposition of forward and backward propagating fields as
\begin{multline}
	{\bf E}^{\pm}(x,y,z)=-ic\iint_{\Omega} [a^{\pm}_{TE}(z) {\bf u}^{\pm}_{TE} + a^{\pm}_{TM}(z) {\bf u}^{\pm}_{TM}]\\ e^{in_{0}k_{0}(s_{x}x+s_{y}y)} \,ds_x\,ds_y
\end{multline}
Here, the $+$ and $-$ signs describe forward and backward propagating waves, respectively, ${\bf{a}}$'s are the amplitudes, and $\bf{u}$'s are the unit vectors along the TE and TM components.   

The tilted stratified medium is modeled by solving the Debye–Wolf integral in a rotated coordinate system whose origin coincides with the nominal focus of the lens, and whose $z$ axis is perpendicular to the stratified medium. This is a three step process where: a) we have to rotate the electric field on the reference sphere $\epsilon$ and all other vector quantities by an angle $\theta_{tilt}$. Thus, the field on the reference sphere will be: $\boldsymbol{\epsilon}^{'}=\bf{T}\boldsymbol{\epsilon}$. Here $\bf{T}$ is a $3\times3$ rotational matrix that takes care of both the axis as well as the angle of rotation. b) Next, we apply the matrix formalism to find the fields $\bf{E^{'}}$ at required positions in the rotated coordinate. c) Finally, we find the field components in the lab frame using the transformation $\bf{E}=\bf{T}^{-1}\bf{E}^{'}$.
\subsection{Transverse spin angular momentum}
Having obtained the electric field near the focal region using the formalism mentioned in the previous section, we proceed to find the distribution of spin density {\bf s}, and the corresponding Belinfante spin momentum density (BSMD)  {$\rm\bf{p^s} $}, \cite{bliokh2014extraordinary} defined by
\begin{gather}
{\bf s}=\frac{1}{4\omega}\bf{Im}[\varepsilon{\bf E}^* \times {\bf E} + \mu {\bf H}^* \times {\bf H}] \equiv {\bf s}_{E} + {\bf s}_{H},
\label{eq_sam}
\\
{\bf p}^s=\frac{1}{2} (\nabla \times {\bf s})
\end{gather}
Here, $\bf{s}_{E} \propto Im[{\bf E}^* \times {\bf E}]$ is the electric contribution and $\bf{s}_{H} \propto Im[{\bf H}^* \times {\bf H}]$ is the magnetic contribution to the total SAM. In general, the intrinsic dispersion of a medium is taken care of by the permittivity ($\varepsilon$) and permeability ($\mu$) of that medium. But for the sake of simplicity, all the layers in the stratified medium are taken to be nondispersive. In a lossless (i.e. refractive index $n$ is real) and nonmagnetic medium ($\mu=\mu_{0}$), the SAM density takes the form: 
\begin{gather}
    \bf{s}=\frac{1}{4\omega}\bf{Im}(\varepsilon_{0}n^2{\bf E}^* \times {\bf E}+\mu_{0}{\bf H}^* \times {\bf H}),
\end{gather}
where, $\varepsilon_{0}$ and $\mu_{0}$ are the permittivity and permeability of free space. It is to be noted that the third Maxwell's equation involving the curl of the electric field has been utilized to compute the magnetic field $\bf{H}$ from the electric field $\bf{E}$. 

For a transversely spinning field, the corresponding transverse spin angular momentum (TSAM) components can be expressed as
\begin{gather}
{\bf s}_x=\frac{1}{4\omega} [Im (E^*_{y} E_{z} - E_{y} E^*_{z} )+ Im (H^*_{y} H_{z} - H_{y} H^*_{z} )]
\\
{\bf s}_y=\frac{1}{4\omega} [Im (E_{x} E^*_{z} - E^*_{x} E_{z} )+ Im (H_{x} H^*_{z} - H^*_{x} H_{z} )],
\end{gather}
where the subscripts $x$ and $y$ indicate the two transverse directions. In general, the total SAM [Eq.~\eqref{eq_sam}] depends on both the electric field and the magnetic field \cite{berry2009optical}. Also, all the three components of SAM along three mutually perpendicular directions are related to the 3D stokes vector parameters \cite{setala2002degree}.  
\section{Results \& Discussions}

The mechanism described in the previous section has been utilized in a MATLAB code for simulating the fields of a Gaussian beam impinging obliquely into a tilted three layer stratified medium. The stratified medium we use in our optical tweezers system is described in Fig.~\ref{tilted_diagram}(b). It consists of the following: light from a laser of wavelength $1064$ nm having a Gaussian intensity profile, is incident on the $100\times$ oil-immersion objective of NA $1.4$ followed by (a) an oil layer of thickness around $5 \mu$m and refractive index (RI) $1.516$, (b) a $160$ $\mu$m thick coverslip having refractive index 1.516/1.814 (henceforth referred to as `matched'/`mismatched' conditions, respectively - note that the `matched condition' is typically employed in optical tweezers to minimize spherical aberration effects at the focal spot), and finally (c) a water layer having a refractive index of 1.33 with a depth of $-6$ $\mu$m. 
The positions of the interfaces with respect to the nominal focus are chosen at -166 $\mu$m and -6 $\mu$m respectively. In general, the microscope oil - being a very viscous liquid - gives us the flexibility to tilt the stratified medium up to a certain maximum angle. Beyond this angle, the oil is not able to cover the whole exit pupil of the objective lens, and the above-mentioned method therefore fails. For the simulation purposes, the maximum tilting angle is taken to be $18^{\circ}$. The tilted stratified medium is modeled as a rotation of the stratified medium about the $y$ axis and rotated around $z=0$. 



We now discuss the simulation results based on the theoretical treatise described above. At first, it is important to note that the stratified medium plays a crucial role in modifying the size and shape of the focal spot. The dependency of the focal spot on the RI contrast has been explored earlier in terms of the diattenuation parameter in Ref.~\cite{PhysRevA.87.043823}. In a nutshell, for large RI contrast, radial and axial lobes of high intensity regions are created near the focal region. Also, in our simulation configuration, with the second layer being optically denser with respect to the third layer, there exists a critical angle beyond which the incident light gets totally reflected. The denser the second medium, the smaller the critical angle. For the second layer of RI $1.516$ and $1.814$ - the critical angles are $61.3^{\circ}$ and $47.1^{\circ}$, respectively. 
Thus, for high NA lenses, the RI mismatch governs the amount of light intensity that can be transmitted into the next layers. When the entire stratified medium is placed axis-symmetrically, the total internal reflection also happens in a similar manner. However, in the case of the tilted stratified medium, the region of the interface that comes closer to the lens reflects a larger amount of incident light back into the second layer compared to the other region that is moved away from the lens. Moreover, on exit from the objective, light waves leave the meridional plane of the objective and follow different planes of incidence. This effectively increases the overall aberration to a larger extent, and breaks the axial symmetry of the entire system. Due to these three effects, we observe significant deviations of the various electromagnetic phenomena from that for the initial axis-symmetric (zero-tilt) configuration.  
\begin{figure}[ht]
	\centering
	\includegraphics[width=0.45\textwidth]{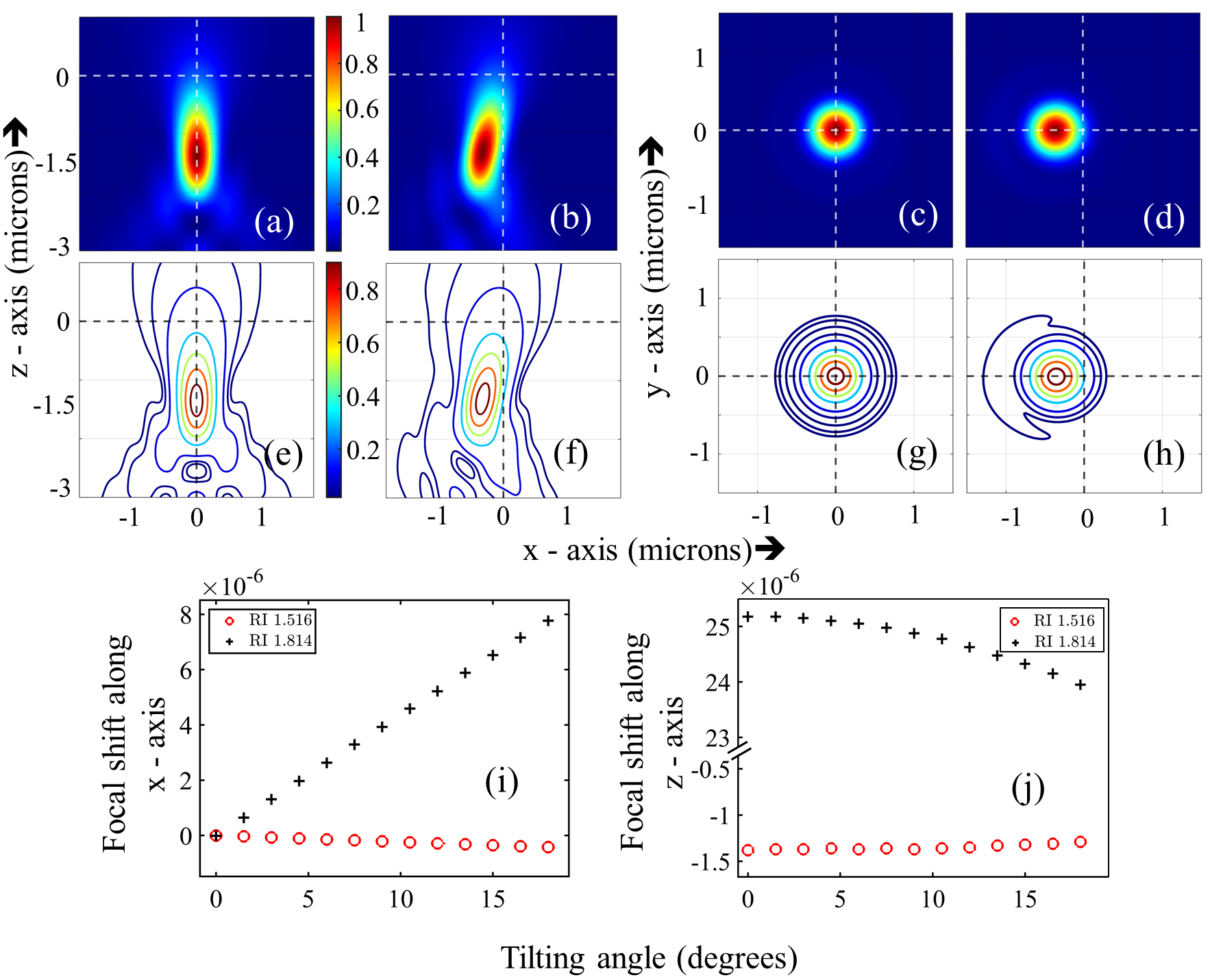}
	\caption{Normalized intensity patterns in the $XZ$ plane for a coverslip RI of 1.516 (a) when there is zero tilt and (b) after applying a tilt of $15^{\circ}$. Normalized intensity pattern in the $XY$ plane for the same coverslip RI at (c) $z=-1.38~ \mu$m and (d) $z=-1.32~ \mu$m with respect to the nominal focus $o$. It is to be noted that $z=-1.38~ \mu$m and $z=-1.32~ \mu$m is the $z$ value of the actual focus when there is zero tilt and at a tilt angle of $15^{\circ}$ respectively. All these intensity patterns correspond to the incident LCP beam. (e-h) The corresponding contour plots in the $XZ$ and $XY$ planes, respectively. Spatial shift of the position of the focus (i) along the $x$ axis and (j) along the $z$ axis with increasing tilt.}
	\label{normintense}
\end{figure}
\begin{figure}[ht]
	\centering
	\includegraphics[width=0.45\textwidth]{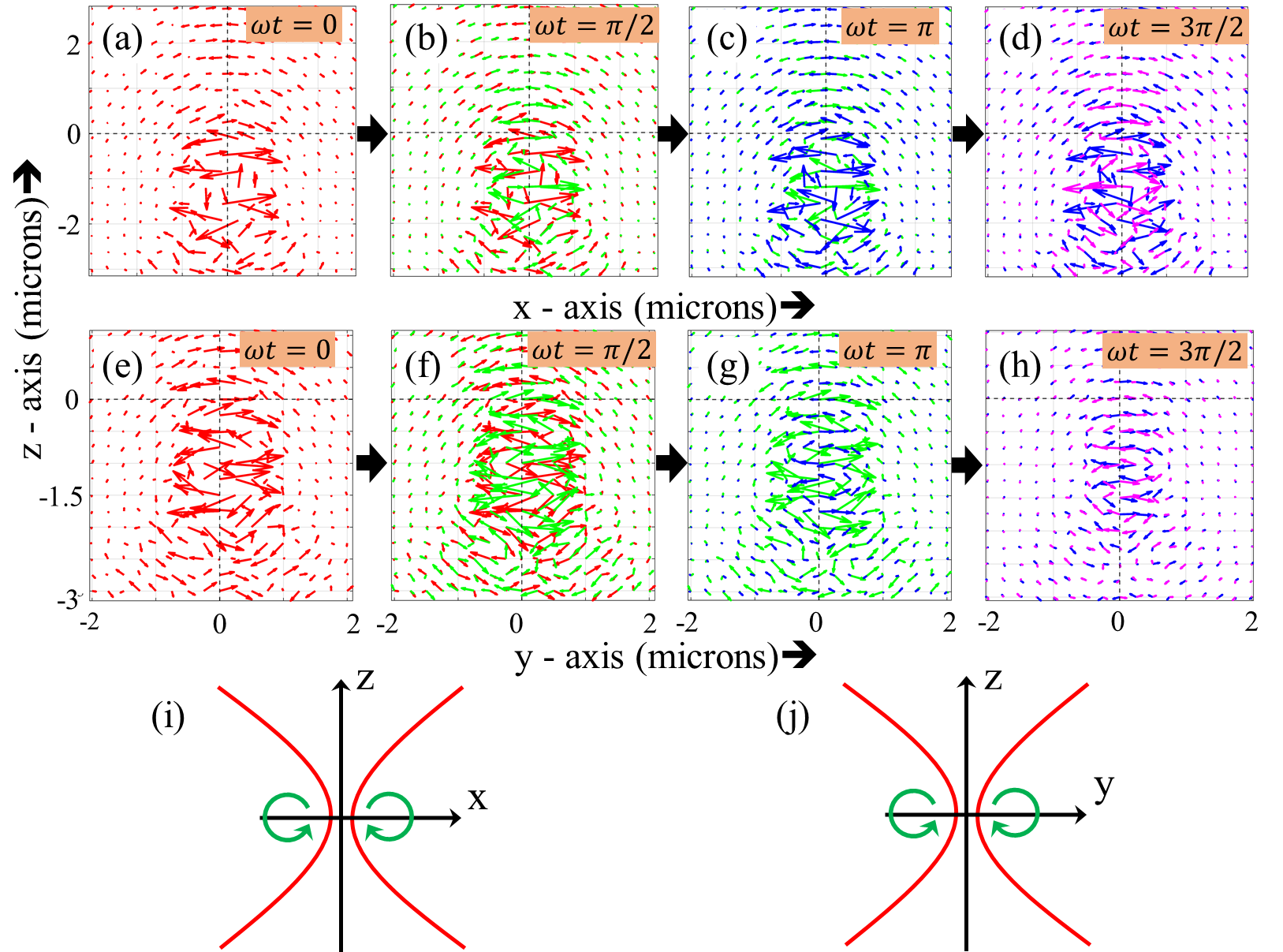}
	\caption{Rotation of the electric field inside the third layer in the $XZ$ ((a) - (d)) , and $YZ$ ((e) - (h)) plane at four consecutive time instants for both RCP and LCP. (i) and (j) are the pictorial representation of the field rotation in the respective planes. The electric field strengths at each point is represented by the corresponding lengths of the arrows. Red, green, blue and purple arrows correspond to $\omega t= 0, \pi/2, \pi, 3\pi/2$, respectively.}
	\label{rotation}
\end{figure}
In order to understand the dependence of TSAM and transverse BSMD generation on tilt and RI contrast of the stratified medium from our simulations, we change the $z$ value of the position of the detector each time the tilt is changed, so as to pick up the maximum value of TSAM from a plane containing the focus. The choice of the plane mainly relies on the spatial shift as well as the shape of the focal spot with tilting as depicted in Fig.~\ref{normintense}(a)-(d). The $XZ$ plane, where the normal to the interface always rotates upon changing the tilt, is referred to as the plane of rotation, which is a common plane of interest at all tilting angles. Note that only the $y$ components of TSAM and BSMD are determined from the $XZ$ plane. In order to find the $x$ components of the TSAM and BSMD, two different longitudinal planes other than the common $XZ$ plane are chosen for the matched and mismatched cases. For the matched case, because of the relatively small focal shift, field components are obtained on the $YZ$ plane for zero tilt, and on planes parallel to the $YZ$ plane for non-zero tilt. Each of these parallel $YZ$ planes always passes through the intensity maximum point of the focal spot at the respective tilt angles. However, since the mismatched case leads to a relatively large focal spot comprising of several intensity maxima lobes accompanied by a large focal shift, the $YZ$ plane is chosen for for zero tilt, while rotated $YZ$ planes are chosen for non-zero tilt. Importantly, these rotated $YZ$ planes always pass through the maximum intensity point of the focus in a similar manner as the parallel $YZ$ planes for the matched case. 

\subsection{Study of the intensity pattern with tilt angle}
One of the immediate consequences of tilting the stratified medium is clear from the intensity distributions around the focal region of the tweezers [see Fig.~\ref{normintense}]. The intensity patterns in the $XY$ and $XZ$ plane, for a particular RI configuration, appear identical at all tilting angles for incident LCP and RCP Gaussian beams - note that we show only the results for an input LCP beam in Fig.~\ref{normintense}. As mentioned earlier, the position of the focal spot experiences a spatial shift with increasing tilt from the zero tilt case as we demonstrate in Fig.~\ref{normintense}(a) and (b) for the $XZ$ plane, and (c) and (d) for the $XY$ plane. For the $XY$ plane, we consider cross-sections at $z=-1.38~\mu$m and $z=-1.32~\mu$m with respect to the nominal focal plane at $z=0$, respectively, for the non-tilted and tilted cases. The intensity contour cases for all these cases are provided in Figs.~\ref{normintense}(e)-(h). Now, the spatial shift along the $z$ axis for the maximum tilt angle ($\theta_{tilt} =18^{\circ}$) for the matched and mismatched coverslips are $0.09 \mu$m and $1.2 \mu$m, respectively, whereas the  shift along the $x$ axis are $0.42 \mu$m and $7.78 \mu$m, respectively. The larger spatial shift in the $XY$ plane compared to the $XZ$ plane occurs since the tilt to the interface is applied along the latter plane, and with the beam being focused at a greater $z$ distance compared to $x$ and $y$ with respect to the origin, the transverse shift accumulated is more than the longitudinal shift. Additionally, the higher RI for the mismatched condition is accompanied by increased spherical aberration \cite{PhysRevA.87.043823}, so that there is a larger transverse shift for the mismatched case compared to the matched case (Fig.~\ref{normintense})(i). In addition, we also observe a considerable smearing out of the focal spot in the axial direction for the mismatched condition due to the increased spherical aberration (Fig.~\ref{normintense})(j). The focal shift, as well as the focal profile play a crucial role in the evolution of the other physics that occur due to the tilting of an interface.
\begin{figure}[ht]
	\centering
	\includegraphics[width=0.45\textwidth]{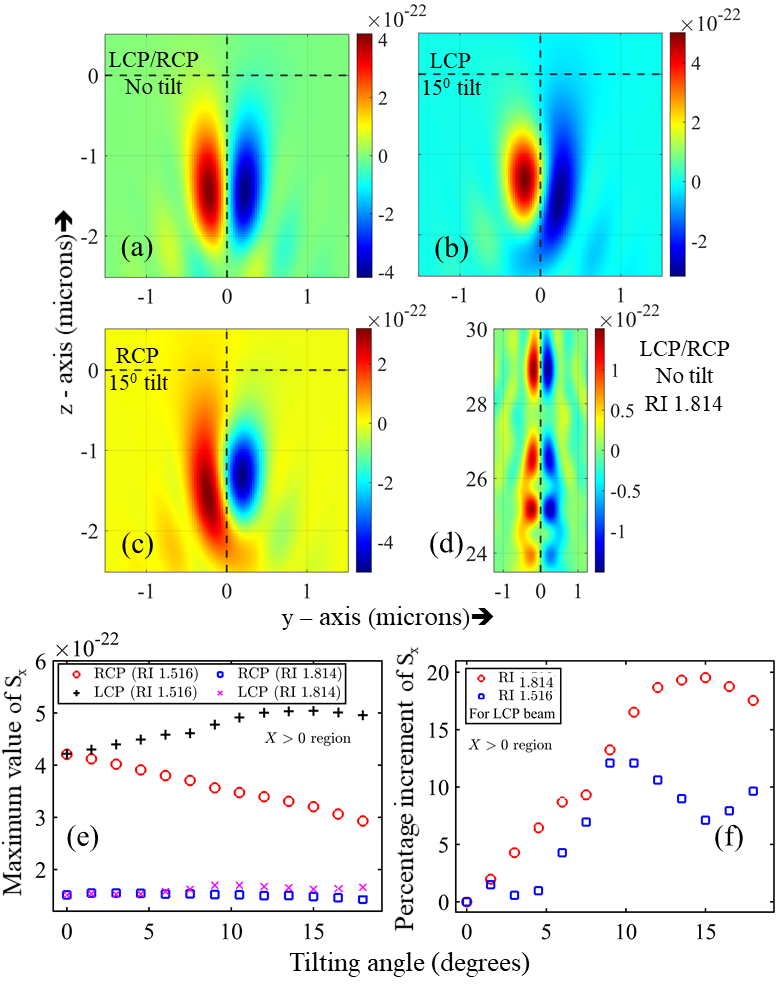}
	\caption{(a) Distribution of $s_{x}$ for the incident LCP beam in the $YZ$ plane for a coverslip RI of 1.516 before applying any tilt in the stratified medium. Without tilt, the distribution of $s_{x}$ for the incident RCP beam in the $YZ$ plane is identical to that for LCP. Distribution of $s_{x}$ for the incident (b) LCP and (c) RCP beams in a plane passing through the focus and parallel to the $YZ$ plane after applying a tilt of $15^{\circ}$ in the matched stratified medium. (d) Distribution of $s_{x}$ density across the focal region for the mismatched case. (e) Angular dependency of maximum value of $s_{x}$  in the positive and negative $X$ regions for the incident RCP beam. For an LCP beam, the variation of $s_{x}$ with tilt angle in the positive and negative $X$ regions is just opposite to that for the LCP beam. (f) Percentage increment of $s_{x}$ with increasing tilt.}
	\label{sam_yz}
\end{figure}
\begin{figure}[ht]
	\centering
	\includegraphics[width=0.45\textwidth]{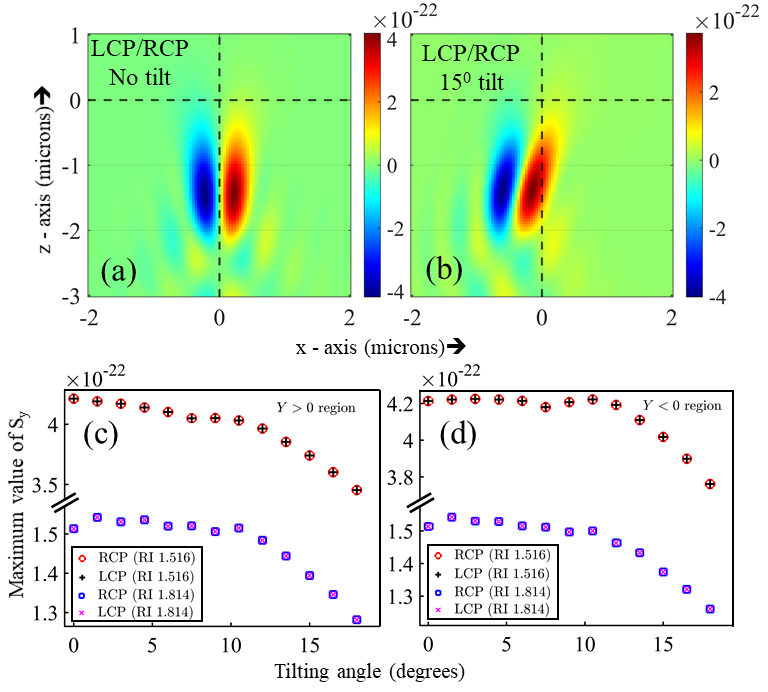}
	\caption{(a) Distribution of $s_y$ in the $XZ$ plane for a coverslip of RI 1.516 before applying any tilt in the stratified medium for RCP/LCP at an wavelength if $1064$ nm.  (b) Distribution of $s_{y}$ for the same coverslip after applying a tilt of $15^{\circ}$. The maximum value of $s_{y}$ in the $XZ$ plane at each tilt angle (c) in the positive $Y$ region and (d) in the negative $Y$ region respectively.}
	\label{sam_xz}
\end{figure}

\subsection{Study of TSAM with tilt angle}

In the course of focusing with a high NA objective lens, the bending of each plane wave generates a longitudinal component of the electric field in a frame attached to the lens at the nominal focus. The phase difference between this induced longitudinal component and the transverse components of the electric field  governs its rotation in the longitudinal planes $XZ$ and $YZ$, which we demonstrate in Figs.~ \ref{rotation}[(a)-(h)]. Indeed, it is also this rotation of the field that is responsible for the generation of transverse SAM (TSAM) near the focal region. It has been shown earlier that the fields in the $XZ$ and $YZ$ planes rotate in the same sense for both LCP and RCP, which clearly describes that the TSAM is helicity independent - a manifestation of spin-momentum locking \cite{pal2020direct}. Our simulations also verify this as is apparent from Fig.~\ref{rotation} (i), where it is clear that in the $XZ$ plane, the electric field rotates clockwise in the positive $X$ region and anticlockwise in the negative $X$ region. Similarly, in the $YZ$ plane [Fig.~\ref{rotation} (j)], clockwise rotation is observed in the positive $Y$ region, and anticlockwise rotation in the the negative $Y$ region. As a result, TSAM is always directed anticlockwise as seen from a point beyond the actual focus [Fig. \ref{tilted_diagram} (a)], and is independent of the input helicity.  For ease of understanding, we pictorial represent the direction of rotation of the TSAM in the $XY$ and $XZ$ planes in Figs.~\ref{rotation}(i) and (j), respectively.

We now investigate the dependence of the TSAM on the angle of tilt of the coverslip. Following our method of detection, we find that [see Figs.~\ref{sam_yz}, \ref{sam_xz}] initially, $s_x$ and $s_y$ have the same maximum value in the positive and the negative regions separated by the $YZ$ and $XZ$ planes. This value remains the same for  both incident LCP and RCP beams. However, the degeneracy breaks as the tilt angle is increased.

Now, from Eq.~(\ref{theta_phi_dependency}), we observe that the field on the reference sphere heavily depends on the azimuthal angle $\phi$. Due to the tilted interface, each plane wave suffers a change in $\phi$ eventually picking up an extra phase factor and contributing significantly differently in the total field (obtained after the double integration is performed) at a point due to the angular dependency of the amount of reflection and transmission from an interface. In general, both the components of TSAM are affected by the action of tilting, but the axis of rotation determines the component that is affected more than the other. We now describe these in more detail.     

\textbf{x component of TSAM:} Fig.~\ref{sam_yz}$(a)$ shows that, without tilt, there are two regions of positive and negative $s_x$ density in the plane of rotation (YZ plane) for both LCP and RCP. However, when tilt is increased, the negative region for input LCP (Fig.~\ref{sam_yz}$(b)$), and the positive region for input RCP  (Fig.~\ref{sam_yz}$(c)$) get smeared out. More interestingly, for the mismatched condition - for zero tilt - we observe a series of axially separated lobes of both positive and negative $s_x$ in (Fig.~\ref{sam_yz}$(d)$. We expect that, for increasing tilt - a similar smearing would occur for the positive and negative regions of the different lobes for input RCP/LCP. We now proceed to quantify the evolution of $s_x$ as a function of tilt angle for the matched and mismatched conditions. For this, we determine the maximum value of $s_x$, for each tilt angle. Thus, as tilt is gradually increased [Fig.~\ref{sam_yz} (e)], $s_x$ for input LCP and RCP in the two half regions starts deviating. While the maximum value of $s_x$ for RCP is observed to increase in the negative $X$ region, that for LCP increases in the positive $X$ region. This increment occurs up to $15^{\circ}$ for the matched case and $10.5^{\circ}$ for the mismatched case, and in the same manner for both input helicity components. In the case of the matched coverslip, beyond $15^{\circ}$ (modified to $10.5^{\circ}$ for the mismatched coverslip), we observe a gradual decrease of the maximum value of $s_x$ up to $18^{\circ}$ for both input LCP and RCP. On the contrary, in the opposite regions, i.e. positive $x$ region for LCP and negative $x$ region for RCP, we observe a constant rate of decrease throughout the entire course of tilting from $0$ to $18^{\circ}$. Thus, there occurs close to $20\%$ increase in the maximum value of $s_x$ at $15^{\circ}$ tilt for the matched case, and a $12\%$ increase for the mismatched case at $10.5^{\circ}$ tilt. The dominance of LCP in the positive $X$ region and RCP in the negative $X$ region up to a certain tilt angle is a manifestation of the broken axial symmetry, where two distinct regions of high TSAM density are observed, depending upon the handedness of the incident light beam.  

\textbf{y component of TSAM:} Fig. \ref{sam_xz}$(a)$ shows that without tilt, there are two regions of positive and negative $s_y$ density in the plane of rotation (XZ plane) for both LCP and RCP. However unlike $s_x$, with tilt, $s_y$ for LCP and RCP behaves in the same way near the focal region in both the longitudinal planes. In the positive $Y$ direction, $s_{y}$ decreases monotonically as a function of the tilt angle. But in the negative $Y$, $s_y$ almost retains its value up to $10.5^{\circ}$, and then starts falling gradually with the tilt angle. 

It is important to note that in the simulation, we provided the tilt about the $y$-axis, as a result of which,  the component of TSAM perpendicular to that axis - i.e. $s_{x}$ - is affected much more than the other component ($s_{y}$). Note that providing the tilt around an arbitrary  axis would result in controlling the TSAM density in a direction transverse to that axis using an appropriate handedness of the incident beam. This feature can be easily exploited in experiments where high TSAM density is required in a particular direction. Another interesting fact regarding the tight focusing of a circularly polarized light is the presence of both the transverse components of TSAM ($x$ and $y$), which is in sharp contrast to the case of evanescent waves and surface plasmons, where it has been shown previously that only one of the TSAM components is generated \cite{saha2016transverse}. Clearly, the presence of both components of TSAM in tight focusing and their tuning by tilting the stratified medium makes this route more interesting and suitable for applications involving the generation of TSAM controllably.

\begin{figure}[ht]
\centering
\includegraphics[width=0.45\textwidth]{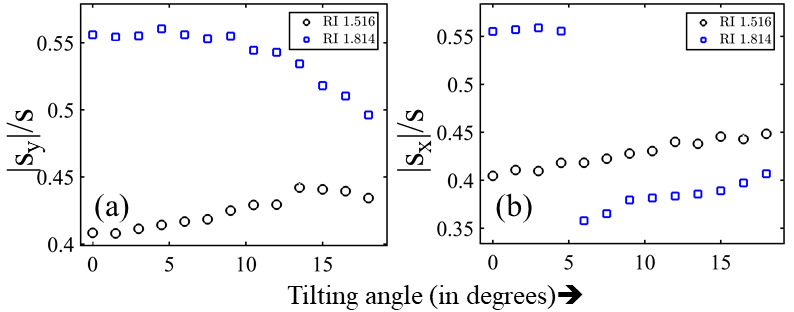}
\caption{(a) Ratio of the maximum values of $s_{y}$ to the total SAM at points where maximum $s_{y}$ are observed in the $XZ$ plane with increasing tilt. (b)  Ratio of the maximum values of $s_{x}$ to the total SAM at points where maximum $s_{x}$ are observed in planes always passing through the focus and parallel to the $YZ$ plane.}
\label{proportion}
\end{figure}

The other interesting question is regarding the extent of conversion of the longitudinal field components into the  transverse  components. Qualitatively, this can be understood from the following argument: any tilt of the plane of incidence by an angle $\theta$ - even for paraxial beams - reduces the LSAM by a factor of $\cos (\theta)$, so that TSAM correspondingly increases since the total SAM of the system is always conserved. For tight focusing, the angle of incidence for each $k$-vector is different, so that once again the LSAM decreases. In this case, though, we have a distribution of LSAM for different $k$-vectors, whose limit is defined by the highest angle of incidence corresponding to the NA of the focusing lens. The TSAM correspondingly increases, as is well known in the literature \cite{bliokh2015physreports, neugebauer2015measuring}. Adding a further tilt angle to the plane of incidence in the tight focusing case increases the maximum angle of the $k$-vectors - so that the TSAM is enhanced even more - as we find out in our simulations. We proceed to study this quantitatively from the ratio of individual components of the TSAM to the total SAM. As mentioned earlier, the ratios are studied at the locations where the maximum value of the individual TSAM components has been observed [see Fig.~\ref{proportion}]. We observe that the ratio of the maximum value of $s_{y}$ to the total SAM behaves differently for the matched and mismatched cases. While this ratio for the mismatched case always decreases as the tilt is increased [blue open squares in Fig.~\ref{proportion}(a)], for the matched case it first increases up to $13.5^{\circ}$ and then decreases slightly [black open circles in Fig. \ref{proportion}(a)].  The ratio of the maximum value of $s_{x}$ to the total SAM, on the other hand, always increases throughout all the tilt angles for the matched case [black open circles in Fig.~\ref{proportion}(b)]. 

The mismatched case is more interesting. As we described earlier, the presence of several high TSAM density lobes in the focal region, leads to the position of the maximum $s_{x}$ shifting from one lobe to another [Fig.~\ref{sam_yz}(d)] with increasing tilt. We observe this to be happening beyond a tilt angle of $4.5^{\circ}$. As a result, there occurs a sharp drop in the ratio of the maximum value of $s_{x}$ to total SAM [blue open squares in Fig.~\ref{proportion}(b)], after which the ratio increases monotonically till the maximum tilt angle. What is interesting, however, is that for both $s_y$ and $s_y$, the TSAM conversion is higher for the mismatched case compared to the matched one. For $s_x$, even though there is a sudden reduction in the TSAM conversion beyond a certain angle due to the presence of several high  density TSAM lobes, those could be useful in experiments since birefringent particles may actually be optically confined and the effects of TSAM observed in them due to the finite physical dimension of the lobes.

\begin{figure}[ht]
\centering
\includegraphics[width=0.45\textwidth]{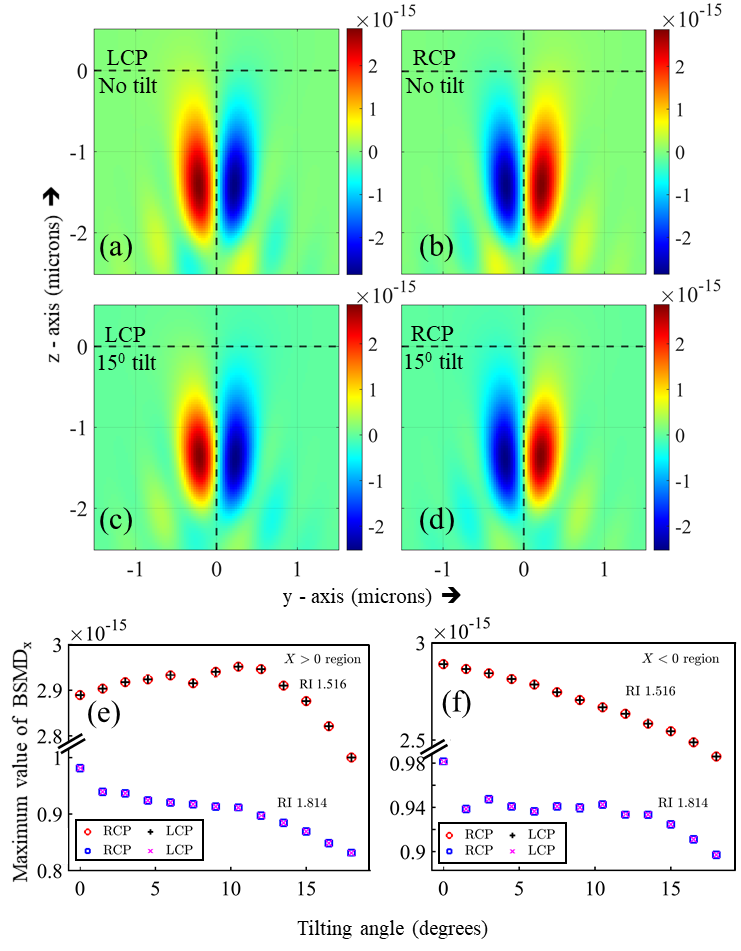}
\caption{Distribution of $\text{BSMD}_{x}$ in the $YZ$ plane for the incident (a) LCP and (b) RCP beam for a coverslip or RI 1.516 before applying any tilt. Distribution of $\text{BSMD}_{x}$ for the same coverslip for (e) LCP and (f) RCP in a plane passing through the focus and parallel to the $YZ$ plane after applying a tilt of $15^{\circ}$. Angular dependency of the maximum value of $\text{BSMD}_{x}$ (c) in the positive $X$ region and (d) in the negative $X$ region for both LCP and RCP.}
\label{bs_yz}
\end{figure}
\subsection{Study of the transverse Belinfante spin angular momentum with tilt angle}
\begin{figure}[ht]
\centering
\includegraphics[width=0.46\textwidth]{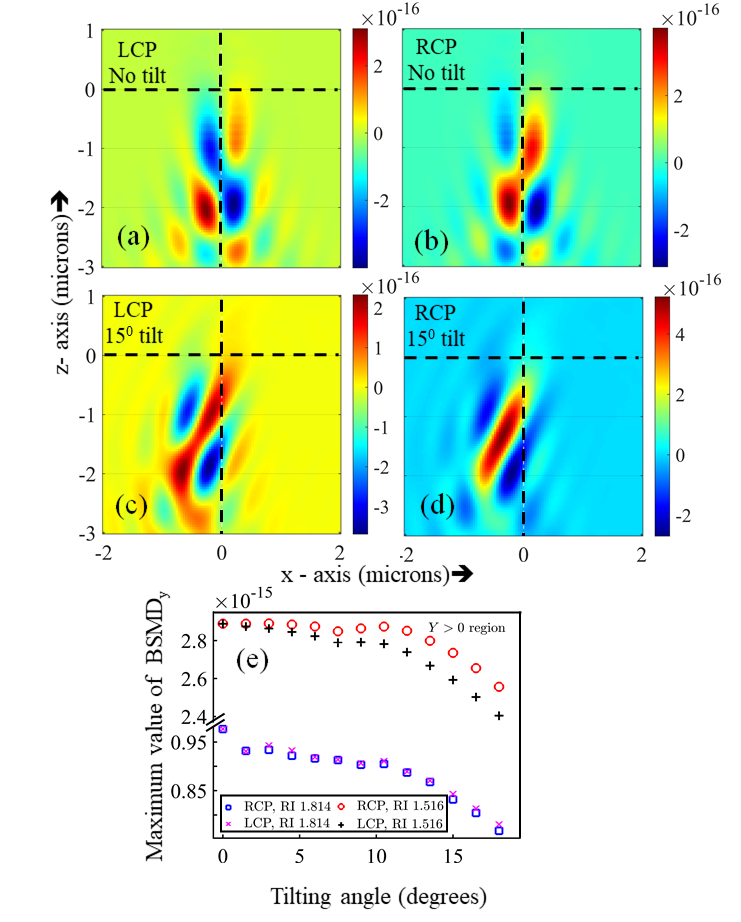}
\caption{Distribution of $\text{BSMD}_{y}$ for the incident LCP and RCP beam of wavelength $1064$nm in the $XZ$ plane for a coverslip or RI 1.516 (a, b) before applying any tilt and (c, d) after applying a tilt of $15^{\circ}$. (e) The maximum value of $\text{BSMD}_{y}$ in the positive $Y$ region at each tilt angle for the incident LCP/RCP beam. In the negative $Y$ region the maximum values of $\text{BSMD}_{y}$ for LCP and RCP get flipped.}
\label{bs_xz}
\end{figure}

The distributions of $\text{BSMD}_x$ and $\text{BSMD}_y$ in the $XZ$ and $YZ$ plane before tilting the stratified medium are displayed in  Fig.~\ref{bs_yz} and \ref{bs_xz}, respectively. In both the planes, we observe spatially separated lobes corresponding to the positive and negative values of $\text{BSMD}_x$ or $\text{BSMD}_y$ depending upon the handedness of the beam. For the incident LCP beam and with zero tilt, $\text{BSMD}_x$ is positive in the negative $Y$ region and negative in the positive $Y$ region. However, for the incident RCP beam, we observe the same spatially separated lobes, but  with opposite values.  The maximum values of $\text{BSMD}_x$ at each tilt angle are found again by moving the detector. In the positive $X$ region, $\text{BSMD}_x$ for both LCP and RCP increases by around $2\%$ at $10.5^{\circ}$ tilt and then starts falling for the matched case. For the mismatched case, we observe a general decrease in $\text{BSMD}_x$ as tilt is increased. In the negative $X$ region, $\text{BSMD}_x$ for both input helicities decreases monotonically, while for the mismatched case we observe a sharp fall when the tilt is applied initially, after which - there appears little change up to an angle of around $10^{\circ}$, before the value starts decreasing. For $\text{BSMD}_y$, there appears little change in the positive $Y$ region up to an angle of around $10^{\circ}$, for input RCP after which the value starts reducing. In this case, however, we again observe a symmetry breaking between input RCP and LCP cases, with the maximum value of $\text{BSMD}_y$ in the positive $y$ region for incident RCP  dominating over that for incident LCP at all tilt angles. However, in the negative $Y$ region, the maximum value of $\text{BSMD}_y$ for LCP dominates over that for the RCP at all tilt angles (simulation results not shown). We also observe the  $\text{BSMD}_x$ lobes remaining separated with increasing tilt. However, the lobes with positive $\text{BSMD}_y$ in the positive $Y$ region overlap with each other.  
\section{Conclusions}
We study - both analytically and by simulations - the evolution of TSAM and transverse BSMD in optical tweezers due to the presence of an RI stratified optical interface that is tilted with respect to the axis of the tightly focused input circularly polarized beam.  The TSAM generated in this case has two orthogonal components, unlike that obtained for evanescent waves of surface plasmons, where only one component is generated. We determine the dependencies of TSAM on tilt angle and the RI contrast of the stratified medium, and identify clear strategies to enhance the LSAM to TSAM conversion by tuning both. We also determine whether a tilt angle exists where the conversion ratio is the maximum for a certain RI. Thus, the TSAM/SAM ratio in the direction perpendicular to the applied tilt ($s_x$) increases as a function of tilt angle for the RI matched case in the positive $X$ direction, while the RI mismatched case is more complex since the intensity distribution itself is more complicated here due to large spherical aberration. Note that the conversion is finally determined by the conservation of the total SAM of the system. The TSAM in the direction  of the applied tilt ($s_y$) generally reduces with the tilt angle for both RI values we consider. For the  BSMD, in the $X$ direction,  we observe an optimum angle of tilt where it is maximised, while in the $Y$ direction, it falls after remaining constant up to some tilt angles.  

We have also observed a symmetry breaking phenomena for circularly polarized symmetric Gaussian beams having opposite helicity when we increase the tilt angle. Thus, $s_{x}$ - one of the orthogonal components of the TSAM - in the YZ plane, and $\text{BSMD}_{y}$ in the XZ plane manifest this breaking of symmetry. Usually, the electric field and the magnetic field for a highly symmetric circularly polarized plane wave contributes equally to the total SAM i.e. $\bf{s}_{E}=\bf{s}_{H}$. This dual symmetry, or electromagnetic democracy, can also be seen for the longitudinal SAM density near the focal region of a tightly focused linearly polarized Gaussian beam (simulation not shown here). The spatially separated areas with high longitudinal SAM density carry equally weighted contributions of ${\bf s }_{z}$ into ${\bf s}_{e_{z}}$ and ${\bf s}_{h_{z}}$ \cite{roy2014manifestations}. However, with the increase of asymmetry - which is brought about by changing the tilt angle - this dual symmetric nature of SAM in the transverse directions does not hold anymore.

The TSAM, together with BSMD of an electromagnetic field, can provide clear insights into understanding the field itself, and thus manipulating it so as to pave a way towards rapid technological advancements in the future using light.  In the present paper, our recipe to tune the TSAM can be extremely useful in optical tweezers to induce complex motion in trapped mesoscopic particles. To understand how the TSAM would affect the dynamics of particles in optical tweezers, it is necessary to determine the torque generated on the particles, for which the Maxwell stress tensor needs to be evaluated for this system. In addition, the SOI induced to tilting will also inevitably lead to the generation of a strong orbital angular momentum (OAM) component, whose properties will also be modified with RI stratification. Indeed, the evolution of OAM in this system, and its manifestations on trapped particles near the focal region of an optical trap will be a fascinating study - both in terms of theory and experiments. We intend to pursue these directions of research in our future work. 
\section{Acknowledgments}
The authors acknowledge the SERB, Department of Science and Technology, Government of India (Project No. EMR/2017/001456). Sauvik Roy is thankful to the Department of Science and Technology (DST), Government of India for INSPIRE fellowship.

\bibliographystyle{apsrev4-2}
\providecommand{\noopsort}[1]{}\providecommand{\singleletter}[1]{#1}%

\end{document}